\documentclass[aps,prl,twocolumn,amsmath,amssymb,amsfonts,nofootinbib,superscriptaddress,altaffilletter]{revtex4-1}
\usepackage{graphicx}
\usepackage{hyperref}
\hypersetup{
  bookmarksopen=true
}
\usepackage{ulem}
\usepackage{amssymb}
\usepackage{amsmath}
\usepackage{color}
\usepackage{supertabular}
\usepackage{dcolumn}

\newcommand{\F}{\mathcal{F}}

\def\sci#1#2{#1\times10^{#2}}

\def\RAJ{\textrm{RA}_{\textrm J2000}}
\def\DECJ{\textrm{DEC}_{\textrm J2000}}

\begin{document}

\title{Results from the first all-sky search for continuous gravitational waves from small-ellipticity sources}

\author{Vladimir Dergachev}
\email{vladimir.dergachev@aei.mpg.de}
\affiliation{Max Planck Institute for Gravitational Physics (Albert Einstein Institute), Callinstrasse 38, 30167 Hannover, Germany}
\affiliation{Leibniz Universit\"at Hannover, D-30167 Hannover, Germany}

\author{Maria Alessandra Papa}
\email{maria.alessandra.papa@aei.mpg.de}
\affiliation{Max Planck Institute for Gravitational Physics (Albert Einstein Institute), Callinstrasse 38, 30167 Hannover, Germany}
\affiliation{Leibniz Universit\"at Hannover, D-30167 Hannover, Germany}
\affiliation{University of Wisconsin Milwaukee, 3135 N Maryland Ave, Milwaukee, WI 53211, USA}

\begin{abstract}
We present the results of an all-sky search for continuous gravitational wave signals with frequencies in the 500-1700\,Hz range targeting neutron stars with ellipticity of $10^{-8}$. The search is done on LIGO O2 data using the Falcon analysis pipeline. 
The results presented here double the sensitivity over any other result on the same data \cite{lvc_O2_allsky,Palomba:2019vxe}. The search is capable of detecting low ellipticity sources up to $170$\,pc.
We establish strict upper limits which hold for worst-case signal parameters. We list outliers uncovered by the search, including several which we cannot associate with any known instrumental cause.
\end{abstract}

\maketitle

{\bf{Introduction}} Continuous gravitational waves are expected over a broad range of frequencies from rapidly rotating compact stars such as neutron stars due to a variety of mechanisms \cite{Lasky:2015uia}, as well as from more exotic scenarios \cite{Brito:2017zvb,boson1,boson2,boson3,Horowitz:2019aim,Horowitz:2019pru}.

A spinning neutron star that presents a deviation from axi-symmetry emits continuous gravitational waves at twice its rotation frequency. Its gravitational-wave brightness is determined by its equatorial ellipticity, which describes how deformed the star is in the plane perpendicular to the rotation axis. Because neutron stars have solar sized mass packed within a tight radius of 10-15\,km, the gravity at their surface is extremely high, making it difficult to support large deformations.

Simulations show that ellipticities as high as $10^{-6}$ can be supported by the crust \cite{crust_limit, crust_limit2}, but at present no mechanism is known  that mandates the existence of such deformations. Searches so far  have failed to find signals from high ellipticity sources \cite{O1AllSky2, allsky3, allsky4, EHO1, lvc_O2_allsky}.

\begin{figure}[htbp]
\includegraphics[width=3.3in]{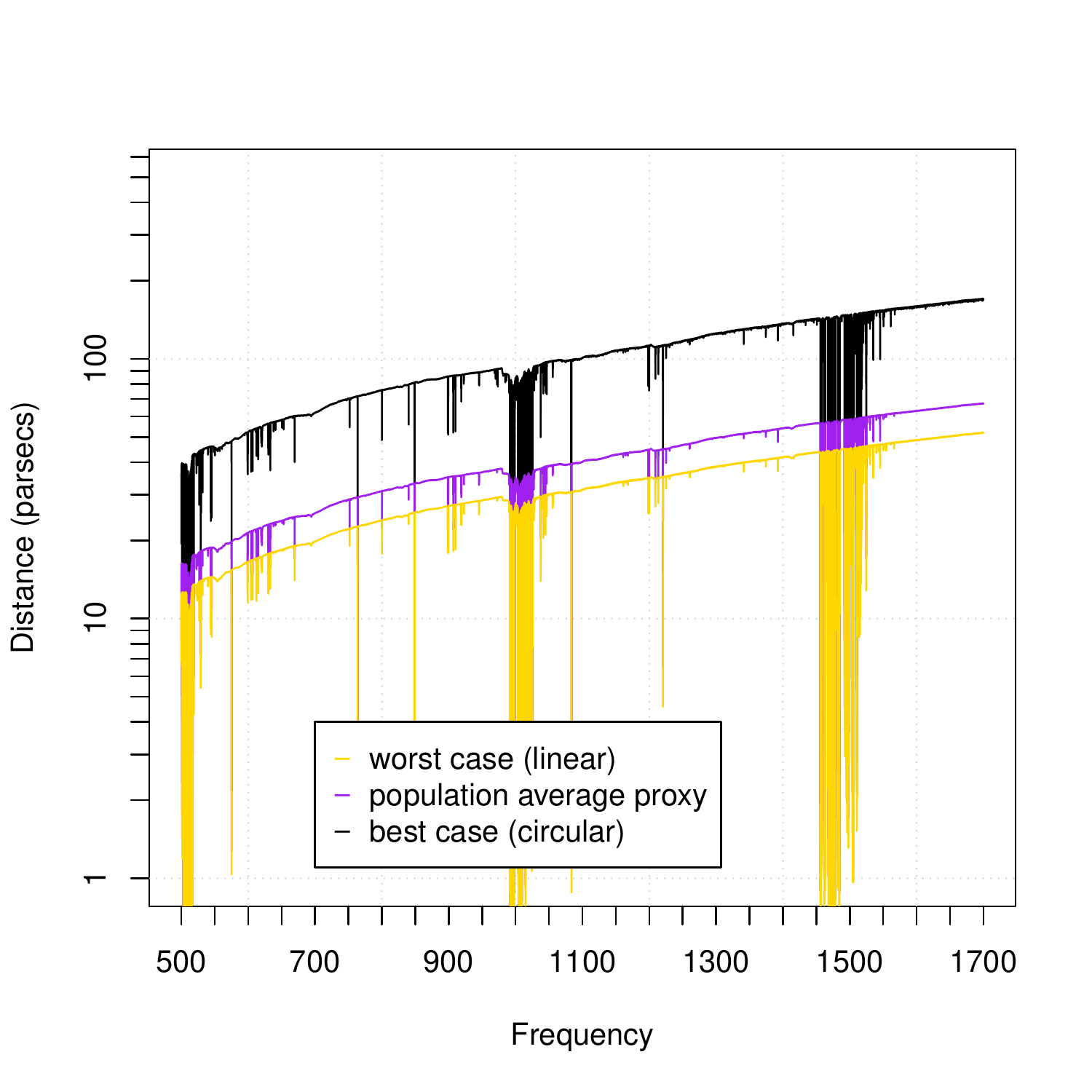}
\caption[Spindown range]{
\label{fig:distance}
Range of the search for stars with ellipticity of $10^{-8}$. The X axis shows gravitational wave frequency, which is twice the pulsar rotation frequency for emission due to an equatorial ellipticity. R-modes and other emission mechanisms give rise to emission at different frequencies.}
\end{figure}

On the other hand, properties of observed pulsars suggest that a population might exist whose spin evolution is governed by gravitational wave emission, with typical ellipticities in the range of $\approx 10^{-9}$ to $10^{-8}$ \cite{ellipticity}. 
This is the range targeted by this search (Figure \ref{fig:distance}). 

We find several outliers and investigate them. We use longer coherence analyses and out-of-sample data. One outlier increases in SNR with increase in coherence length, however, this increase is not as large as one would expect for an ideal signal on average.

The basic signal assumed at the gravitational wave detector is a nearly monochromatic signal from an isolated source (i.e. not in a binary), with Doppler modulations due to the relative motion between the source and the detector and amplitude modulations due to the relative orientation of the detector to the source. Optimal sensitivity is typically achieved with a large degree of phase coherence in the signal and for signals matching the assumed waveform model.

No past search has revealed any such signal, over a broad range of possible signal parameters. One reason could be that past searches were not sensitive enough and have not yet probed low enough values of the ellipticity. Another reason (or perhaps an additional reason)  might be that the signals deviate enough from the assumed model that the search methods cannot identify them with sufficient confidence. 

Deviations from the naive signal model could come about in a number of ways, for example due to the motion of the source being affected by another body, or the signal itself not being perfectly stable. While standard searches might retain very good sensitivity to many of these signals \cite{Singh:2019han}, upon closer inspection a signal candidate might be disregarded because it does not pass all the tests that the naive signal would pass.

As more and more sensitive searches yield null results it seems wise to entertain the possibility that the signals are not quite as we assume. Here we do this by highlighting not only the upper limit results for the standard signals, but also the top outliers that are marginal/inconclusive candidates. This will enable cross-checks with data that we have no access to, and may lead to the identification of an instrumental or environmental artefact, or a signal.

{\bf{The method}} Broad-band all-sky searches for continuous gravitational waves are computationally challenging. The most sensitive surveys employ multi-stage searches that rely on clever search methods, computationally efficient algorithms and much computing power. 

A key characteristic of long-duration searches is the employed coherence length. 
Increasing the coherence length increases the search sensitivity and makes the search more selective, being able to distinguish more finely different signal parameters like sky location and frequency. The number of separate points of parameter space that need to be explored grows with the increasing coherence length and this generates greater computational demands. The computational cost can scale as the fourth power of the coherence length for even simple searches and it grows very quickly 
for searches taking into account more complex signals, with higher order frequency derivatives or in the presence of a companion object in a binary system.

A common way to construct a search with a given coherence length is to partition the data into segments spanning that coherence length, perform a matched filter analysis on each piece and then combine the results. This is known as a ``semi-coherent'' search, because the results of the coherent matched filter analysis are combined with no phase information. See \cite{allsky3, EHO1, Walsh:2019nmr, Walsh:2016hyc, depths} and references therein for an overview and comparison of the different families of semi-coherent search methods.

Loosely coherent methods \cite{loosely_coherent, loosely_coherent2, loosely_coherent3} are constructed by partitioning the parameter space of the searched signals into families of closely related signals that can be considered to be perturbations of each other. The search is then designed to work on each family separately 
and because many signals are searched for ``at the same time'' it leverages significant economy of scale. 

\begin{figure}[htbp]
\begin{center}
   \includegraphics[width=3.6in]{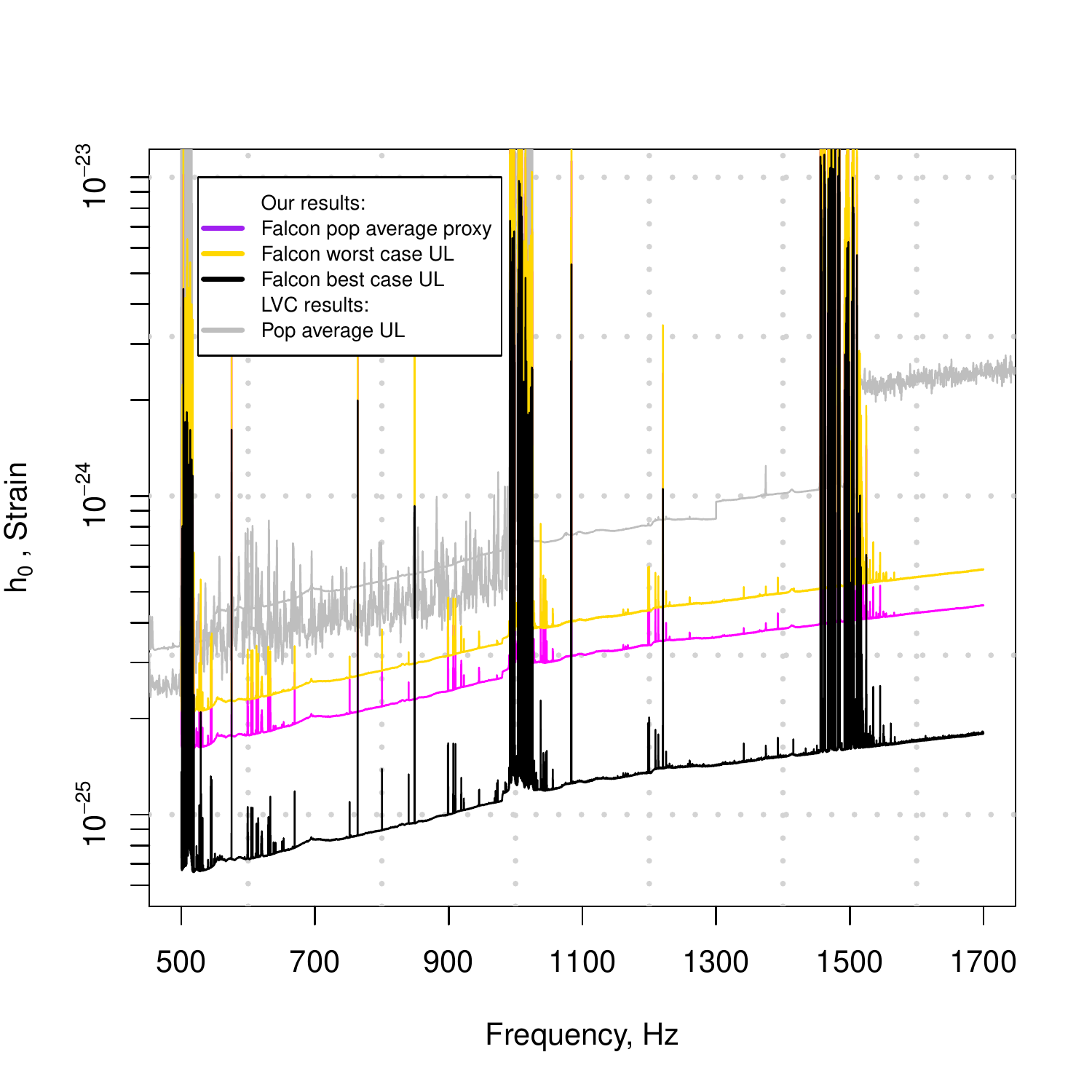}
 \caption{The intrinsic gravitational-wave strain-amplitude upper-limit (vertical axis) is plotted against signal frequency. 
 The lower curve (black) shows the best-case upper limits (circularly polarized signals), the next curve (purple) shows the population-average proxy, then (in yellow) we have the worst-case upper limits. The fainter lines at the top are the upper limits from the latest LIGO-Virgo collaboration (LVC) all-sky survey \cite{lvc_O2_allsky} and they can be directly compared to our population-average ones (middle purple curve). The LVC results are produced independently by three different pipelines which cover overlapping frequency ranges.}
\label{fig:O2_upper_limits}
\end{center}
\end{figure}

The newest loosely coherent implementation targeting longer coherent timescales, the Falcon search, represents a breakthrough in performance and sensitivity \cite{allsky3}. We have explored the LIGO O1-run data employing the Falcon search with a first-stage coherence length of 4 hours, and demonstrated its performance and computational efficiency \cite{allsky3,allsky4}. In this paper we push the envelope further, doubling the size of the input data, tripling the coherence length of the first stage, extending the frequency range by a factor of nearly three and searching high frequencies\footnote{The necessary computing power scales as the cube of highest frequency searched}. We attain more than a two-fold increase in strain sensitivity with respect to the best LIGO all-sky search on the same data (Figure \ref{fig:O2_upper_limits}).

{\bf{The search}} The computational load of an all-sky  search per-Hz-searched is massively different in the kHz region and in the $O(100)$\,Hz region. This means that in order to achieve optimal sensitivity different search set-ups should be used in different frequency ranges (this argument is clearly illustrated in \cite{Ming:2015jla,Ming:2017anf}). We choose here to tackle the challenging high-frequency region. We aim at detecting continuous gravitational wave signals with frequency between  $500-1700$\,Hz and first-order frequency derivative $|f_1 | \leq \sci{3}{-12}$\,Hz/s for frequencies below $1000$\,Hz and $\leq \sci{2.5}{-12}$\,Hz/s above $1000$\,Hz. The spindown range is consistent with our ellipticity target at a level $< 10^{-8}$. It is also generous: the largest known frequency drift among pulsars spinning faster than $250$\,Hz  is $-\sci{1.73}{-13}$\,Hz/s, for J1824-2452A spinning at $327$\,Hz  \cite{ATNF}. The full search parameter range is given in Table \ref{tab:parameter_space}.

We use the public LIGO \cite{aligo} O2 data \cite{o2_data, losc}. Below 1000 Hz this data was subject to a cleaning procedure  that removed a substantial amount of spurious instrumental noise \cite{O2_cleaning}. In general this procedure will contribute an additional systematic uncertainty to the calibration, which is however not stated. We use the available O2 data as is, thus our results below $1000$\,Hz have a potential systematic uncertainty that overall should match the uncertainty (5\% and 10\% for H1 and L1 respectively) given in LIGO papers such as \cite{lvc_O2_allsky}.

The search begins with a coherence length of 12\,hours that is extended to 6\,days through 3 follow-up stages (Table \ref{tab:pipeline_parameters}). The coherence length increases from stage to stage and with it the search sensitivity also increases. Candidates pass to the next stage only if their detection statistic value exceeds a predetermined threshold value, which increases from one stage to the next. The last stage also includes consistency checks between the single-instrument results. Candidates that survive this entire sequence of tests represent one of the results of the search. We refer to these as the {\it {outliers}} from the search. Every outlier is defined by a set of signal parameters (frequency, sky position and frequency derivative) and signal to noise ratio value (SNR).

\begin{table}[htbp]
\begin{center}
\begin{tabular}{rD{.}{.}{2}D{.}{.}{3}}\hline
Stage & \multicolumn{1}{c}{Coherence length (hours)} & \multicolumn{1}{c}{Minimum SNR}\\
\hline
\hline
0  & 12 & 6 \\
1  & 24 & 7 \\
2  & 48 & 8 \\
3  & 144 & 16 \\
\hline
\end{tabular}
\end{center}
\caption{Parameters for each stage of the search. The stage 3 outliers are subject to an additional consistency check.}
\label{tab:pipeline_parameters}
\end{table}
\begin{table}[htbp]
\begin{center}
\begin{tabular}{lcc}
\hline
\hline
\rule{0pt}{3ex} 
Band & $ | f_1 |_{\textrm{max}}$  & $|f_2|_{\textrm{max}} $\\
\scriptsize{$~~~$[Hz]} & \scriptsize{{$10^{-12}$ [Hz/s]}} & \scriptsize{{$10^{-20}$ [Hz/s$^2$]}}\\
\hline
\hline
\rule{0pt}{3ex}
500-1000  & 3 & 4 \\
1000-1700  & 2.5 & 2 \\
\hline
\end{tabular}
\end{center}
\caption{Signal-parameter space covered in this search. Both positive and negative values of first and second order frequency derivatives $f_1$ and $f_2$ are explored.}
\label{tab:parameter_space}
\end{table}

{\bf{Upper limit results}}
The other results of this paper are the 95\% confidence-level continuous gravitational wave amplitude upper limits $h_0$. These represent the smallest amplitude of a continuous signal with a given frequency, coming from an arbitrary direction in the sky and with frequency drift in our search range,
that we can exclude from impinging on the detectors. 
We have assumed a quasi-monochromatic signal with slow evolution in frequency that can be approximated by a quadratic model:
\begin{equation}
f(t)=f_0+(t-t_0)f_1+(t-t_0)^2f_2/2,
\label{eq:freqEvolution}
\end{equation}
where $f_0$ is the signal frequency at GPS epoch $t_0=1183375935$, and $f_1$ and $f_2$ are the linear and quadratic frequency drifts, respectively.

Our upper limits are established based on the observed power estimates and are valid even in the most heavily contaminated spectral regions. We use a procedure based on the so-called ``universal statistics''  \cite{universal_statistics} which determines the smallest gravitational wave intrinsic signal amplitude $h_0$ consistent with the observed power. This produces valid upper limits without assumptions on the shape of the noise distribution. By construction the upper limits also hold in the presence of a signal. 

We compute three upper-limits: the so-called worst-case upper limit which is the largest upper limit over those established for individual sky locations, polarization\footnote{maximum over polarization is usually reached for linear polarizations}, spindown and frequency; the circular-polarization upper limit which is the largest upper limit over those established for individual sky locations, spindown and frequency, having assumed circular polarization of the signal; and finally a proxy for the population-average upper limit which is determined over a population of sources uniformly distributed on the sky and with all possible polarizations (uniform in the cosine of the inclination angle ($\cos\iota$) and polarization angle $\psi$). This last upper limit is provided 
for ease of comparison with other search results  
\cite{O1LowFreq, O1AllSky2, EHO1, Palomba:2019vxe} and it is estimated (hence it is a proxy) as a weighted average of the upper limits from the individual polarizations. 

The population average proxy is verified by directly measuring the detection efficiency of fake signals added to the real data with frequencies in the  sample bands 500-600\,Hz and 1075.40-1100\,Hz band. 
Each fake signal has an intrinsic strain amplitude $h_0$ equal to the population-average upper-limit proxy value from its frequency band;  other parameters are uniformly distributed in the searched parameter space, apart for the spindown values that are log-uniformly distributed in the search range. 
The full analysis is performed for each fake signal, covering the entire sky and including outlier follow-up. A fake signal is considered detected when an outlier is found within $\sci{5}{-5}$\,Hz of the signal frequency $f$, within $10^{-12}$\,Hz/s of its spindown, and within $0.06\textrm{\,Hz}/f$ radians of its sky location, the latter calculated after projecting on the ecliptic plane (``ecliptic distance'').
We obtain 95\% recovery of injections 
everywhere except the heavily contaminated violin mode region,  and even in those bands the detection efficiency does not drop lower than 90\%.

The upper limits are plotted in Figure \ref{fig:O2_upper_limits}. The upper limit data is available in computer-readable format at \cite{data}. 

Assuming an optimally oriented source of ellipticity $10^{-8}$ we are sensitive as far as $170$\,pc at $1700$\,Hz. Figure \ref{fig:distance} shows the reach of the search to such sources. Above $700$\,Hz, with the exception of a few contaminated frequency bands, we can exclude the presence of such sources (aside from the outliers listed in Table \ref{tab:Outliers}) within a $20$\,pc distance from Earth. 

It is reasonable to expect neutron stars from our target population within this range: assuming a galactic population of $10^9$ neutron stars, if the density in space of neutron stars is the same as that of the known pulsars, then within 20 pc of Earth we expect\footnote{This conservative estimate assumes spherical distribution of neutron stars. Since 1.36\,kpc is large compared to the thickness of Milky Way disk and size of galactic arms, we expect that the true number is much larger.} to have as many neutron stars and we observe as pulsars in $\sim 1.36$\,kpc. At the time of writing the ATNF catalog shows 56 pulsars rotating above 250 Hz within 1.36\,kpc of Earth and of these 12 are isolated.

Our upper limits are also relevant for boson condensates around black holes \cite{boson1,boson2}, which are expected to emit monochromatic continuous wave signals \cite{discovering_axions}. Indeed, if ultralight bosons exist, they will form clouds around rotating black holes via superradiance instability and through annihilations and level transitions will emit continuous gravitational waves. We leave it to the interested reader to constrain from our upper limits physical quantities of interest, based on the specific model they wish to consider. Assuming the ensemble signal of \cite{Zhu:2020tht} from a galactic population of $O(10^8)$ isolated stellar mass black holes with maximum mass 30 $M_\odot$ and maximum initial spin uniformly distributed in  [0,1], our results exclude bosons with masses in the range $\approx [1.0 - 3.0] \times 10^{-12}$ eV. 

{\bf{Outliers}} The full list of outliers is available in \cite{data}. Table  \ref{tab:Outliers} shows a summary of this list produced by displaying the largest SNR outlier in every 0.1\,Hz band. This is especially helpful for the loud fake signals present in the data stream for validation purposes \cite{hardwareInjectionsURL}, since these produce many outliers. Because these fake signals are produced directly at the LIGO detectors by an appropriate excitation of its mirrors, they are commonly referred to as ``hardware injections''. In Table \ref{tab:injections} for convenience we give the parameters of these fake signals.

Our top 8 outliers in Table \ref{tab:Outliers} are caused by such hardware-injections. The large majority of the remaining outliers are due to large hardware artifacts, as indicated in the last column of \ref{tab:Outliers}. There remain outliers whose origin could perhaps be identified based on data from physical and environmental monitoring channels. Since this data is not public we cannot perform such investigations. 

A key question is whether any of the listed outliers are produced by an astrophysical source. The simulations performed to verify the population average proxy assure us that if a signal were present in the data with detectable strength, there would be a corresponding outlier in Table \ref{tab:Outliers}. 
Clearly none of the outliers due to hardware injections are astrophysical. A number of outliers are not due to hardware injected signals but are located in frequency bands with instrumental contamination. As the pipeline is perfectly able to detect signals in most heavily contaminated bands this does not rule out these outliers. However, it does cast doubt on their significance as contamination can artificially increase the SNR in a single or both instruments.

\begin{figure}[htbp]
\includegraphics[width=3.3in]{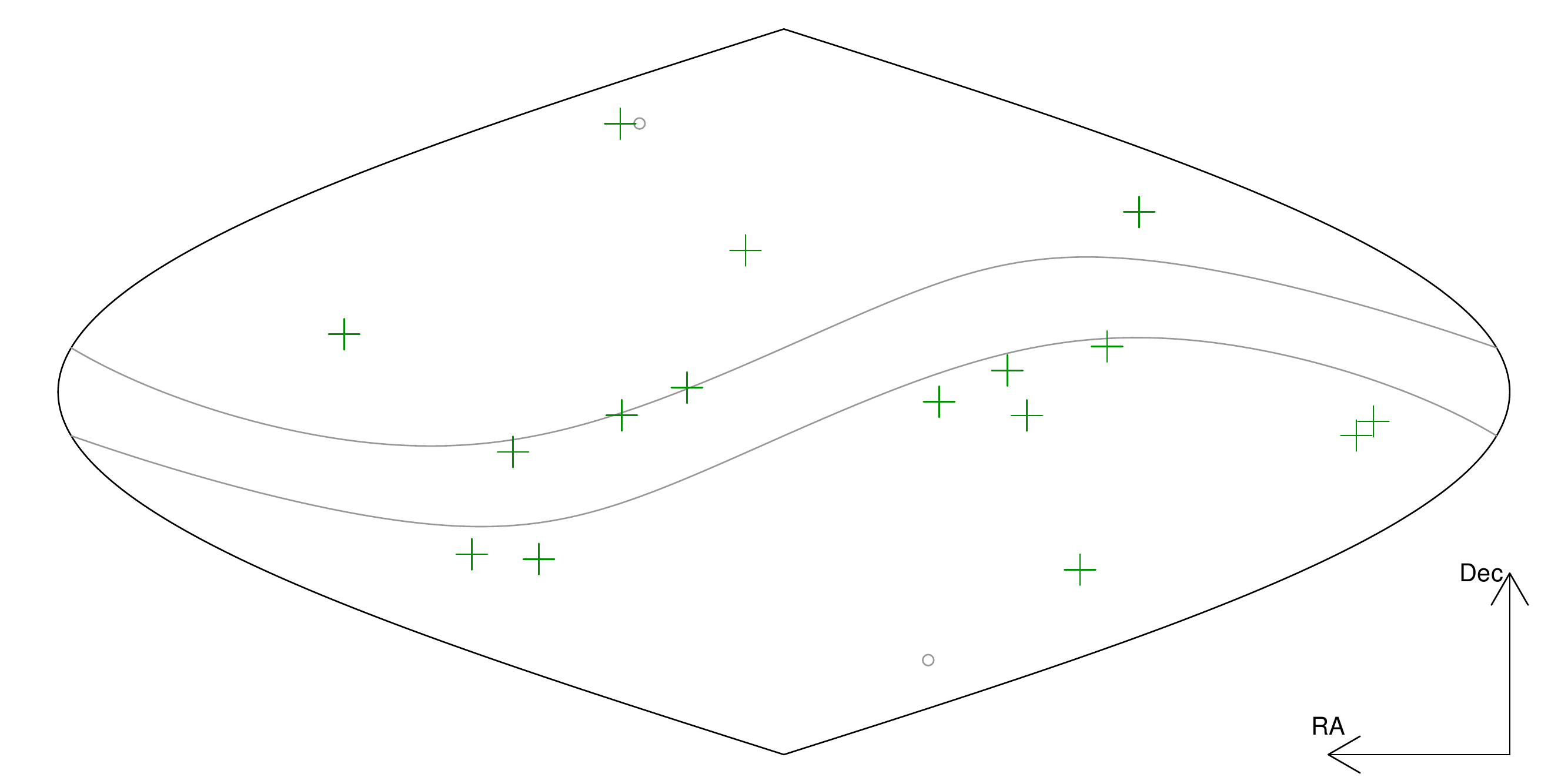}
\caption[Outlier map]{
\label{fig:outlier_map}
Location in the sky of the outliers not associated to any instrumental disturbance. The wide band shows the ecliptic plane, the two small circles are the ecliptic poles.
}
\end{figure}

The 16 outliers not associated with any known instrumental disturbance are the most interesting. Their sky locations are shown in Figure \ref{fig:outlier_map}. 
They do not appear to cluster in any location tied to instrumental artifacts, such as the ecliptic pole. If the outliers were due to astrophysical signals that follow the frequency model of Eq.~\ref{eq:freqEvolution}, then the uncertainty on their true location would be within the limits established by the Monte Carlo simulations discussed above, i.e. $0.06\textrm{\,Hz}/f$ radians, projected on the ecliptic plane.

\begin{figure}[htbp]
\includegraphics[width=3.3in]{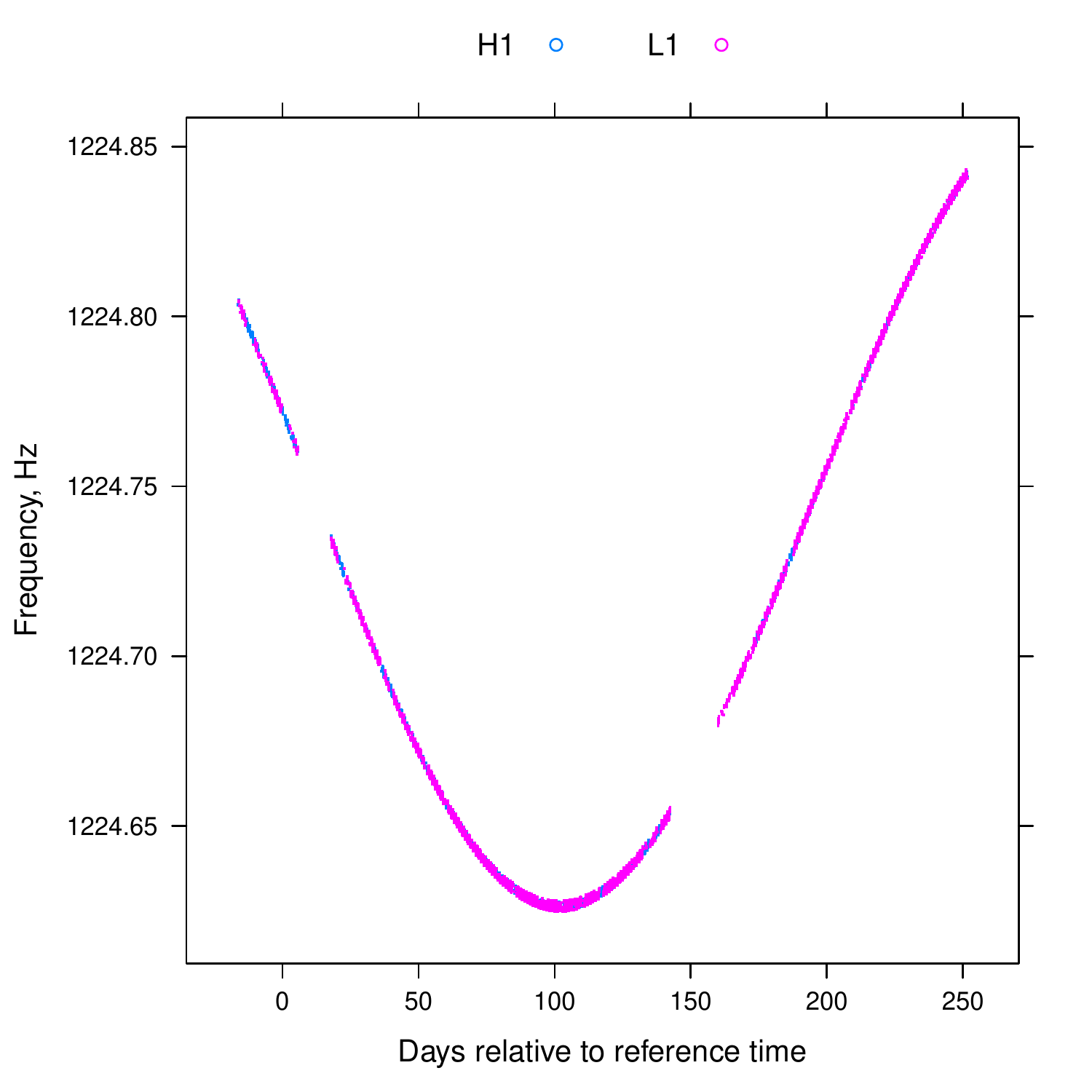}
\caption[Outlier map]{
\label{fig:outlier_evolution}
Apparent frequency of a signal with parameters equal to those of outlier 43 at 1224.74\,Hz, at the detectors. The difference in Doppler shifts between interferometers is small compared to the Doppler shifts from the Earth's orbital motion.
}
\end{figure}

\begin{figure}[htbp]
\includegraphics[width=3.3in]{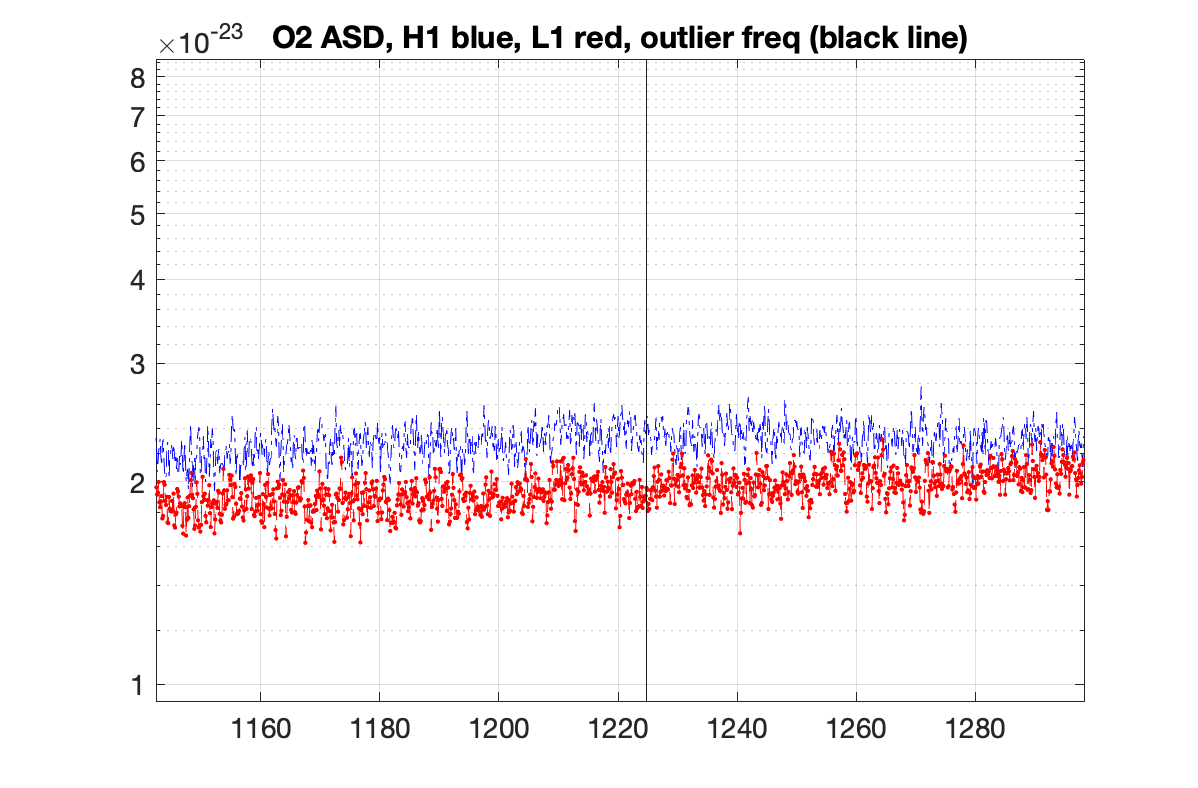}
\caption[Outlier spectrum]{
\label{fig:outlier_hist}
Average amplitude spectral density around the frequency of outlier 43.}
\end{figure}

The frequency spectrum of the data around these outliers is manually examined and none reveals any obvious contamination. As an example we show plots (Figures \ref{fig:outlier_evolution} and \ref{fig:outlier_hist}) corresponding to the highest-frequency outlier 43 at 1224.75\,Hz. 
The apparent frequency of a signal with the parameters of outlier 43, spans almost 0.25\,Hz during the O2 observing time and its evolution is shown in Figure \ref{fig:outlier_evolution}.

The other 15 outliers have similar characteristics. The frequency evolution curves change depending on the outlier parameters but they all exhibit an appreciable modulation. 

Ad-hoc investigations are focused on outliers 25 and 43, which are consistently found with both the Falcon and semi-coherent F-statistic analyses. Follow-ups over longer coherence lengths (up to 500 hours) suggest that the most promising outlier is outlier 25. We note that the $f_1$ value of outlier 25 is positive which would mean that, if the outlier were associated to an astrophysical continuous signal, what is observed is {\it {not}} directly related to the spin frequency of the star.

The SNR build-up for outlier 25 over searches with increasingly long coherent time baseline is not inconsistent with what is expected from a signal. This holds true also for the fully coherent search over the O2 data, albeit the recovered SNR is lower than the expectation value based on the previous stages and with indications that the coherence might not be completely stable \footnote{The most significant measured 2$\F$ value is $\approx$ 47. 2$\F$ is the optimal multi-detector fully coherent detection statistic  \cite{Cutler:2005hc} for signals of the type of Eq.~\ref{eq:freqEvolution}.}. A follow-up on O1 data is inconclusive: the detection statistic value is not inconsistent with the expectations under the signal hypothesis, and the Gaussian-noise p-value is {{$\lesssim 1\%$}}.

As already pointed out, we are unable to investigate the O1 and O2 instrumental data channels, as this data is not part of either the O1 or O2 public data releases.

A first detection can be \'a-la GW150914 (the first gravitational wave signal \cite{Abbott:2016blz}) with a single loud event that spectacularly meets all the predictions; a less glamorous scenario is that evidence builds up over time with outliers first identified in a broad survey like this and later consistently recovered on new data or through the identification of an electromagnetic counterpart. In this spirit we report and discuss the outliers that we find.

Our outliers are not present in the outlier or sub-threshold candidate lists from earlier papers. This is not surprising, because
previous searches on LIGO O2 data that cover the parameter space of this search \cite{lvc_O2_allsky, Palomba:2019vxe} are significantly less sensitive.

The O3 data is roughly a factor of 2 more sensitive than the O2 data employed in this search so the LVC broad-band searches should be able to probe continuous gravitational wave amplitudes comparable to these. Also, targeted searches based on the information presented here, which are comparatively less complex than these broad surveys, should be straightforward.
We look forward to comparing our results on O3 data with those from the LVC, and we are ready to perform fast-turnaround investigations on new data.

{
\begin{table*}[htbp]
\begin{center}
%\scriptsize
%\tiny
\begin{tabular}{D{.}{.}{2}D{.}{.}{3}D{.}{.}{5}D{.}{.}{4}D{.}{.}{4}D{.}{.}{4}l}\hline
\multicolumn{1}{c}{Idx} & \multicolumn{1}{c}{SNR}   &  \multicolumn{1}{c}{Frequency} & \multicolumn{1}{c}{Spindown} &  \multicolumn{1}{c}{$\RAJ$}  & \multicolumn{1}{c}{$\DECJ$} & Description \\
\multicolumn{1}{c}{}	&  \multicolumn{1}{c}{}	&  \multicolumn{1}{c}{Hz}	&  \multicolumn{1}{c}{pHz/s} & \multicolumn{1}{c}{degrees} & \multicolumn{1}{c}{degrees} & \\
\hline \hline
% idx & % snr & % frequency & % spindown & % ra & % dec & % Description
1 & 15792.0 & 763.84732 & -0.20 & 198.894 & 75.690 & Injected pulsar ip9\\
2 & 11970.1 & 575.16352 & -0.20 & 215.256 & 3.445 & Injected pulsar ip2\\
3 & 2470.1 & 848.95675 & -0.80 & 38.563 & -30.432 & Induced by injected pulsar ip1\\
4 & 88.7 & 1220.55117 & 0.00 & 218.862 & 24.074 & Injected pulsar ip7\\
5 & 64.1 & 575.20332 & -1.80 & 184.911 & 38.083 & Induced by injected pulsar ip2\\
6 & 38.1 & 849.00348 & 2.60 & 67.737 & -11.166 & Induced by injected pulsar ip1\\
7 & 36.3 & 848.89699 & 1.00 & 353.084 & -56.879 & Induced by injected pulsar ip1\\
8 & 26.4 & 763.77799 & -2.20 & 222.633 & -68.689 & Induced by injected pulsar ip9\\
9 & 22.3 & 1004.27599 & -1.08 & 294.094 & -88.258 & Disturbed H1 spectrum, lines in L1\\
10 & 19.8 & 986.28829 & -3.20 & 21.424 & 75.556 & Coincident lines in H1 and L1\\
11 & 19.4 & 1006.24309 & 1.08 & 97.733 & -67.657 & Disturbed H1 spectrum, lines in L1\\
12 & 18.4 & 575.09518 & -2.00 & 296.704 & 30.014 & Induced by injected pulsar ip2\\
13 & 18.2 & 990.50363 & -3.60 & 318.140 & 61.085 & Line in L1\\
14 & 18.1 & 1392.54524 & -0.70 & 236.558 & 43.756 & Induced by injected pulsar ip4\\
15 & 17.7 & 900.21880 & -2.20 & 119.428 & -5.859 & \\
16 & 17.6 & 1021.84249 & -0.70 & 13.986 & -8.813 & Sharp line in L1\\
17 & 17.6 & 968.58972 & -1.80 & 338.891 & -76.549 & Line in H1\\
18 & 17.4 & 1392.47501 & -0.72 & 215.544 & -33.472 & Induced by injected pulsar ip4\\
19 & 17.4 & 514.14893 & 1.60 & 124.356 & 5.300 & \\
20 & 16.9 & 1220.62294 & 1.44 & 256.566 & 65.131 & Induced by injected pulsar ip7\\
21 & 16.8 & 997.54586 & -1.80 & 210.349 & 46.174 & Disturbed H1 spectrum\\
22 & 16.8 & 995.20385 & 1.40 & 205.113 & 48.184 & Disturbed H1 spectrum\\
23 & 16.7 & 1001.36623 & 4.30 & 77.684 & -44.133 & \\
24 & 16.6 & 676.19542 & 2.80 & 220.418 & -5.822 & \\
25 & 16.6 & 744.21917 & 2.40 & 191.654 & 35.097 & \\
26 & 16.5 & 998.97311 & -2.00 & 51.221 & 11.566 & Disturbed H1 spectrum, line in L1\\
27 & 16.4 & 997.41470 & 1.40 & 345.963 & -27.957 & Disturbed H1 spectrum\\
28 & 16.4 & 991.29230 & 2.20 & 48.300 & -60.960 & Disturbed H1 spectrum\\
29 & 16.4 & 512.49081 & 1.20 & 141.462 & -2.467 & \\
30 & 16.3 & 991.10725 & 0.40 & 313.415 & 61.938 & Disturbed H1 spectrum, line in L1\\
31 & 16.3 & 983.15189 & 2.20 & 204.037 & 1.030 & \\
32 & 16.3 & 988.60680 & 1.80 & 145.053 & -44.240 & Lines in H1 and L1\\
33 & 16.3 & 510.79524 & -2.40 & 23.196 & -5.131 & Disturbed L1 spectrum\\
34 & 16.2 & 886.88009 & -1.60 & 281.482 & -40.307 & \\
35 & 16.2 & 988.37320 & 1.20 & 56.255 & 44.595 & \\
36 & 16.2 & 512.26233 & 3.00 & 334.102 & -12.737 & Line in L1\\
37 & 16.2 & 963.34281 & -2.80 & 306.751 & 59.163 & Line in H1\\
38 & 16.2 & 1580.37607 & -2.50 & 11.198 & 32.482 & Line in L1\\
39 & 16.1 & 514.29168 & 3.20 & 32.603 & -7.354 & \\
40 & 16.1 & 831.98847 & 0.40 & 281.743 & 66.494 & \\
41 & 16.1 & 873.52461 & 0.40 & 35.466 & -10.855 & \\
42 & 16.1 & 895.42195 & 3.60 & 292.529 & 14.276 & \\
43 & 16.0 & 1224.74567 & -2.16 & 98.278 & 11.241 & \\
44 & 16.0 & 1014.34897 & -2.50 & 25.330 & 82.271 & Lines in H1 and L1\\
45 & 16.0 & 698.72803 & -0.20 & 261.117 & -41.490 & \\
46 & 16.0 & 1095.55740 & -1.08 & 249.504 & -14.911 & \\

\hline
\end{tabular}
\caption[Outliers produced by the detection pipeline]{Outliers produced by the detection pipeline. Only the highest-SNR outlier is shown for each 0.1\,Hz frequency region. Outliers marked with ``line'' have strong narrowband disturbances near the outlier location.
Signal frequencies refer to GPS epoch $1183375935$.} 
\label{tab:Outliers}
\end{center}
\end{table*}
}

\begin{table}[htbp]
\begin{center}
\begin{tabular}{lD{.}{.}{6}rD{.}{.}{5}D{.}{.}{4}}
\hline
Label & \multicolumn{1}{c}{Frequency} & \multicolumn{1}{c}{Spindown} & \multicolumn{1}{c}{$\RAJ$} & \multicolumn{1}{c}{$\DECJ$} \\
 & \multicolumn{1}{c}{Hz} & \multicolumn{1}{c}{pHz/s} & \multicolumn{1}{c}{degrees} & \multicolumn{1}{c}{degrees} \\
\hline \hline
ip1   &  848.969641  & 300   &   37.39385     &  -29.45246 \\
ip2   &  575.163521  & -0.137   &  215.25617     &    3.44399 \\
ip4   & 1393.540559  & -254   &  279.98768     &  -12.4666  \\
ip7   & 1220.555270 & -1120     &  223.42562     &  -20.45063 \\
ip9   &  763.847316 & $\sci{-1.45}{-5}$    &  198.88558     &   75.68959 \\
\hline
\end{tabular}
\caption[Parameters of hardware injections]{Parameters of the hardware-injected simulated continuous wave signals during the O2 data run (GPS epoch $1130529362$).}
\label{tab:injections}
\end{center}
\end{table}

\section{Acknowledgements}
The search was performed on the ATLAS cluster at AEI Hannover. We thank Bruce Allen, Carsten Aulbert and Henning Fehrmann for their support. We also acknowledge useful discussion with Bruce Allen on the distribution of nearby neutron stars.

This research has made use of data, software and/or web tools obtained from the LIGO Open Science Center (\url{https://losc.ligo.org}), a service of LIGO Laboratory, the LIGO Scientific Collaboration and the Virgo Collaboration.  LIGO is funded by the U.S. National Science Foundation. Virgo is funded by the French Centre National de Recherche Scientifique (CNRS), the Italian Istituto Nazionale della Fisica Nucleare (INFN) and the Dutch Nikhef, with contributions by Polish and Hungarian institutes.

\end{document}